# The third conformer of graphane: A first principles DFT based study

AUTHOR NAMES: A. Bhattacharya[1], S. Bhattacharya[1], C. Majumdar[2], G.P. Das[1*]

AUTHOR ADDRESS: [1]Department of Materials Science, Indian Association for the Cultivation of Science, Jadavpur, Kolkata 700032, India,

[2]Chemistry Division, Bhabha Atomic Research Centre, Mumbai 400085, India,

[1*]Department of Materials Science, Indian Association for the Cultivation of Science, Jadavpur, Kolkata. 700032, India.

AUTHOR EMAIL ADDRESS: [1]msab2@iacs.res.in, [1]mssb3@iacs.res.in, [2]chimaju@barc.gov.in, [1*]msgpd@iacs.res.in.

CORRESPONDING AUTHOR FOOTNOTE: [1*]G. P. Das, Department of Materials Science, Indian Association for the Cultivation of Science, Jadavpur, Kolkata-700032, India. Phone No: +91-33-2473 4971 Extn. 202 (Office), Fax: +91-33-2473 2805.

**Abstract:**

We propose, on the basis of our first principles density functional based calculations, a new isomer of graphane, in which the C-H bonds of a hexagon alternate in 3-up, 3-down fashion on either side of the sheet. This 2D puckered structure called 'stirrup' has got a comparable stability with the previously discovered chair and boat conformers of graphane. The physico-chemical properties of this third conformer are found to be similar to the other two conformers of graphane with an insulating direct band gap of 3.1 eV at the $\Gamma$ point. Any other alternative hydrogenation of the graphene sheet disrupts its symmetric puckered geometry and turns out to be energetically less favorable.

**Manuscript Text:**

Graphene[1], the two dimensional (2D) array of hexagonal units of $sp^2$-bonded C atoms, has initiated considerable scientific interest over the past with its unusual electrical and mechanical properties[2]. The nanoribbons and nanoflakes of graphene show high magnetic moment depending upon their edge states[3]. Graphene nanosheet has been used for the storage of molecular hydrogen[4]. Recently, a fully hydrogenated graphene sheet, called graphane, which is an 'extended two dimensional covalently bonded hydrocarbon', having a formula unit CH, was predicted by Sofo et. al[5]. Subsequently, this compound has been synthesized in the laboratory[6] and various interesting properties of this two dimensional sheet, such as reversible hydrogenation-dehydrogenation[6], insulating wide band gap[5,7], magnetization by partial dehydrogenation of the sheet[8] etc have been reported. This hydrocarbon is predicted to have two possible conformers viz. 'chair' and 'boat', having 2D puckered honeycomb like structure with one hydrogen atom bonded covalently ($sp^3$) to each carbon atom of the sheet. In the chair conformer, the H atoms attached to the C atoms of the sheet alternates on both sides of the sheet (1up-1down as in Fig. 1a & 1b) while in the boat conformer, the C bonded H atoms alternates in pair (2up-2down as in Fig. 1c & 1d). It has been shown by Sofo et al. that these conformers of graphane have higher binding energies compared to many known members of the hydrocarbon family (methane, cyclohexane, polyethylene, acetylene and benzene) with an insulating direct band gap at the Γ point. Like graphane, many other synthetic hydrocarbon materials have fascinated the researchers with their unexpected geometries and application based properties. In this context, we would like to mention cubane ($C_8H_8$) having the same C:H ratio as graphane, which is found to be stable despite of its cubic structure with 90º C-C-C bond angles[9]. Although the experimental stability of cubane molecule (4.47 eV/atom) is lower than that of benzene ring (4.82 eV/atom) because of its highly strained C-H bonds, but it still follows a stable kinetics due to unavailability of readily decomposable paths[10]. Thus, the hydrocarbon family is comprised of many bizarre members with peculiar properties. The objective of this brief report is to present our first-principles based prediction of the existence of a new conformer of graphane. We name this new conformer as 'Stirrup', which has stability comparable to the other two conformers of graphane. The structural and physical properties of this conformer have been compared with the other two conformers of graphane and are discussed here in details.

Our calculations have been carried out using first-principles density functional theory (DFT)[11, 12] based on total energy calculations within the generalized gradient approximation (GGA). We have used VASP[13] code with projected augmented wave (PAW) potential[14] for all elemental constituents; viz. H and C. The GGA calculations have been performed using the PW91 exchange correlation functional of Perdew et al[15,16]. An energy cut off of 600eV has been used. The k-mesh was generated by Monkhorst–Pack[17] method and the results were tested for convergence with respect to mesh size. In all our calculations, self-consistency has been achieved with a 0.0001 eV convergence in total energy. For optimizing the ground state geometry[18, 19], atomic forces were converged to less than 0.001eV/Å via conjugate gradient minimization. In order to analyze the stability of the structure we have implemented *ab initio* molecular dynamics simulation (MD) using Nose thermostat[20]. The Mulliken population analysis[21] and vibrational frequency analysis have been carried out by DMOL3 code[22] where the GGA calculations are performed by PW91 exchange-correlation. A double zeta numerical basis set with a polarization function (DNP) has been used.

In the stirrup structure (Fig.1e and 1f), proposed in this report each carbon atom is bonded to a hydrogen atom in such a way that three consecutive H atoms of each hexagon alternate on both sides of the sheet (ie. 3up- 3down), in contrast to the chair or boat conformer where the H atoms alternates singly or pair wise in either side of the C-plane respectively. Fig 1e shows a stirrup hexagonal unit with three H atoms pointing up and three pointing down the C plane (encircled in blue and red respectively). The space group, lattice parameter, atomic positions, bond lengths and the physical properties of all three conformers are given in Table1. In the stirrup conformer, the angles that each H-C bond makes with the three adjoined C atoms of the sheet (∟HCC) are all about 105°, which suggests a tendency towards $sp^3$ hybridization in the bonding between C and H atoms of the sheet. Stirrup has two different C-C bond distances; those connecting C atoms bonded to H atoms lying on the same side of the C-plane, have a bond length of 1.53 Å while the bond which connects two C atoms bonded to H atoms on opposite sides of C-plane are 1.58 Å. The boat conformer of graphane also has two different C-C bond lengths (Fig 1c and 1d). However, unlike the stirrup structure, the C-C bond lengths in the boat conformer are higher for the C atoms bonded to the H atoms on the same side of the plane (Table 1). This is because of the subtle difference in hybridization between the C and H in the boat and stirrup conformer. In the boat conformer, the angles between each H-C bond and the adjoined C atoms of the sheet are about 107°, which indicates higher degree of the sp$^3$ hybridization in the boat than the stirrup conformer. In the chair conformer (Fig 1a and 1b), all C-C bond connects the H atoms attached on opposite sides of C-plane and therefore has only one value of C-C bond length (1.5 Å) through out the sheet (Table1). All the HCC angles in the chair conformer are about 108°. Thus the chair conformer shows higher covalency in the nature of bonding than the boat and stirrup conformers. Thus the difference in hybridization of the three conformers is reflected in difference in the bond length and bond angles of the three structures. However, the C-H bond length is the same (~1.10 Å) for all the three conformers.

The binding energy (BE) of a stirrup structure having 6 C and 6 H atoms in one unit cell is calculated by the formula;

$$BE = \frac{E_{total}(Stirrup) - 6 \times E_{total}(C) - 6 \times E_{total}(H)}{12}$$

We find that a stirrup structure has BE of 5.16 eV/ atom, which is comparable to our estimated binding energy of benzene ring (5.15 eV/ atom), having the same C:H ratio. As obtained from our DFT based calculation, the chair conformer has the highest stability with a BE of 5.21 eV/atom, while the boat conformer is calculated to have a BE of 5.18 eV/atom. Comparing with the Table 1 of Ref. 5, we find that our BE values that we have estimated are consistently lower than the estimated BE values by Sofo et. al. However, the relative trend in magnitude remains unchanged. The interlayer bonding between two stirrup layers is found to be negligible, which is similar to that of the other conformers of graphane. This is a consequence of the saturated bonding between the C and H atoms in the sheet.

The BE of one H atom bonded to the C atom of the sheet for a stirrup structure is estimated to be ~ 4.10 eV, the corresponding values for chair (~ 4.38 eV) and boat (~4.22 eV) conformers are higher. This may favor the reversible hydrogenation-dehydrogenation[7] of a stirrup sheet in the laboratory.

In order to verify the stability of the structure, we have performed 5ps *ab initio* molecular dynamics simulation using Nose algorithm[20] with 1fs time step and 5000 SCF run. Our detailed MD simulation suggests that the stirrup structure is quite stable at very high temperature (~1000K). In Fig 4 we have plotted the average relaxation of the C-C and C-H bonds with the increase in time step as estimated at room temperature (T=300K). The average bond distances do not vary considerably through out the MD run which does suggest that it is possible to realize the stirrup structure at room temperature (300K).

The salient features of the electronic structures of the stirrup conformer can be seen from the band structure in the Γ→M→K plane and the site-projected densities of states (p-DOS), as shown in Fig. 2. As is well known, a pure graphene sheet, the dehydrogenated counter part of graphane, is a semimetal with its valence band and conduction band merging to a Dirac cone at the Fermi level. However, the hydrogenation of the graphene sheet leads to opening of the band gap, there by increasing the inherent stability of the single sheet 2D material. The Stirrup structure is found to have an insulating band gap of 3.1 eV. Although the band gap of the chair (3.5 eV) and boat (3.3 eV) conformers are higher than that of the stirrup conformer, all three have a direct band gap at the Γ point (Fig.2). From the p-DOS plot, hybridization between the H and C atoms of the sheet can be seen in both occupied and unoccupied part of the DOS (Fig.2).

We have also carried out spin polarized calculation of the sheet and found that the spin up and spin down DOS to be identical, thus ruling out any signature of magnetism in the system.

In order to calculate the charge distribution between the C and H atoms of the stirrup conformer, we have deployed the Mulliken population analysis scheme. The stirrup structure shows an electronic charge state of 0.10 for the H atoms while the C atoms have a charge state of -0.10 (Table-1) which is lower than the corresponding charge state in both chair (0.13) and boat (0.15) conformers. The Mulliken population analysis of a benzene ring also shows an electronic charge state similar to the chair conformer of graphane. However, the boat and stirrup conformers undergoes higher and lower charge transfer respectively compared to the chair configuration which leads to the weakening or strengthening of C-C bonds in these two isomeric compounds. In order to show the nature of bonding between the H and C atoms, we have given the orbital and charge density contour plots of the stirrup conformer (Fig.3). From the highest occupied molecular orbital (HOMO) plot of the stirrup conformer, hybridization between the s-orbital of hydrogen and p-orbital of carbon can be seen (Fig.3a). The fig also shows a bonding orbital between the p-orbitals of two neighboring carbon atoms. The charge contour plot of the plane perpendicular to the C-plane (fig.3b) shows a high charge density (violet) in between the C-H and C-C bonds of the stirrup conformer, which confirms a sharing of electrons between the H and C atoms of the sheet. Thus, it suggests a covalent nature of bonding among the atoms of the sheet.

The vibrational frequency analysis has been carried for a 4X4 supercell of the stirrup conformer with 16 Å vacuum. It shows positive frequencies for all the normal modes of vibration of the stirrup conformer, indicating the inherent stability of the structure. The highest vibrational frequency mode of the stirrup conformer corresponds to the C-H stretching mode, and occurs at a frequency of 3122 cm$^{-1}$. This frequency is IR active and is higher than the highest stretching mode vibrational frequency of the chair and boat conformers (Table 1). This is due to the lower binding of the H atoms to the sheet in stirrup than the other two conformers as discussed above.

Any attempt other than these three to increase the number of C-H bonds on one side of the sheet disrupts the parity as well as the periodic puckeredness of the structure. We have carried out some test calculations with 4up-2down or 5up-1down etc. in one hexagon and found that these structures (configurations) are energetically less favorable by 0.13 eV/atom or more.

In summary, we present our first principles DFT based calculations to propose a new conformer of graphane called 'stirrup'. The ground state electronic properties of this new conformer have been compared to the other two conformers of graphane and are found to be similar. This structure has got a comparable stability (albeit marginally less) with the chair and boat conformers of graphane. It is an insulator with a direct energy band gap of 3.1 eV at the $\Gamma$ point. Any other alternative hydrogenation of the graphene sheet disrupts the symmetric puckered geometry of the structure and turns out to be energetically less favorable.

FIGURE CAPTIONS (Color Online):

Fig 1: Ball and stick model of the chair, boat and stirrup conformers, with black balls representing the carbon atoms and white balls representing the hydrogen atoms. The H atoms pointing up and down the C plane are encircled in blue and red respectively. (a), (c) and (e) show hexagonal unit of chair, boat and stirrup conformers respectively. (b), (d) and (f) show the lateral view of chair, boat and stirrup layers respectively.

Fig 2: Band structure and density of state (site-projected) plots of a stirrup conformer.

Fig 3: Lateral view of stirrup structure showing (a) orbital plot of highest occupied molecular orbital (HOMO) (b) charge density contour plot perpendicular to the C- plane.

Fig 4: Molecular dynamics simulation showing the average relaxation of C-H and C-C bonds at 300K with 5ps time period and 1fs time step.

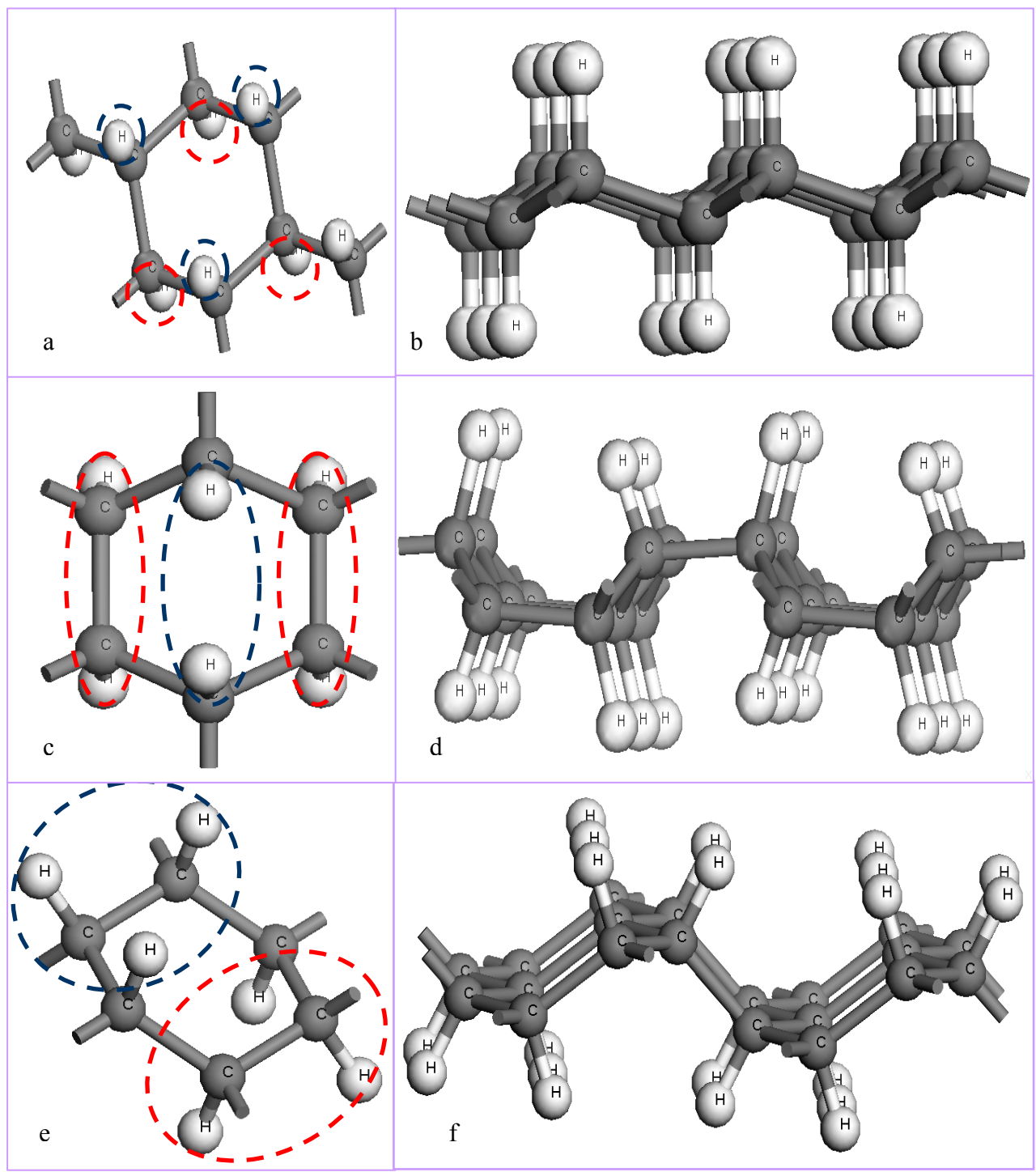

Fig 1.

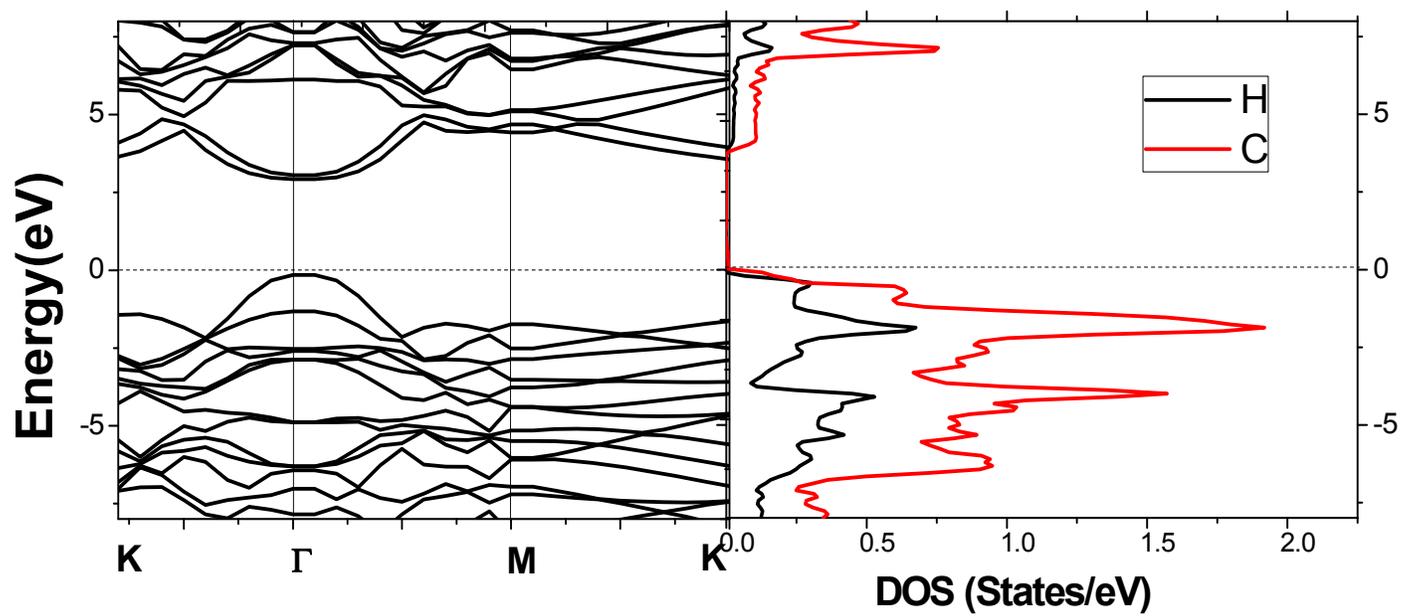

Fig 2.

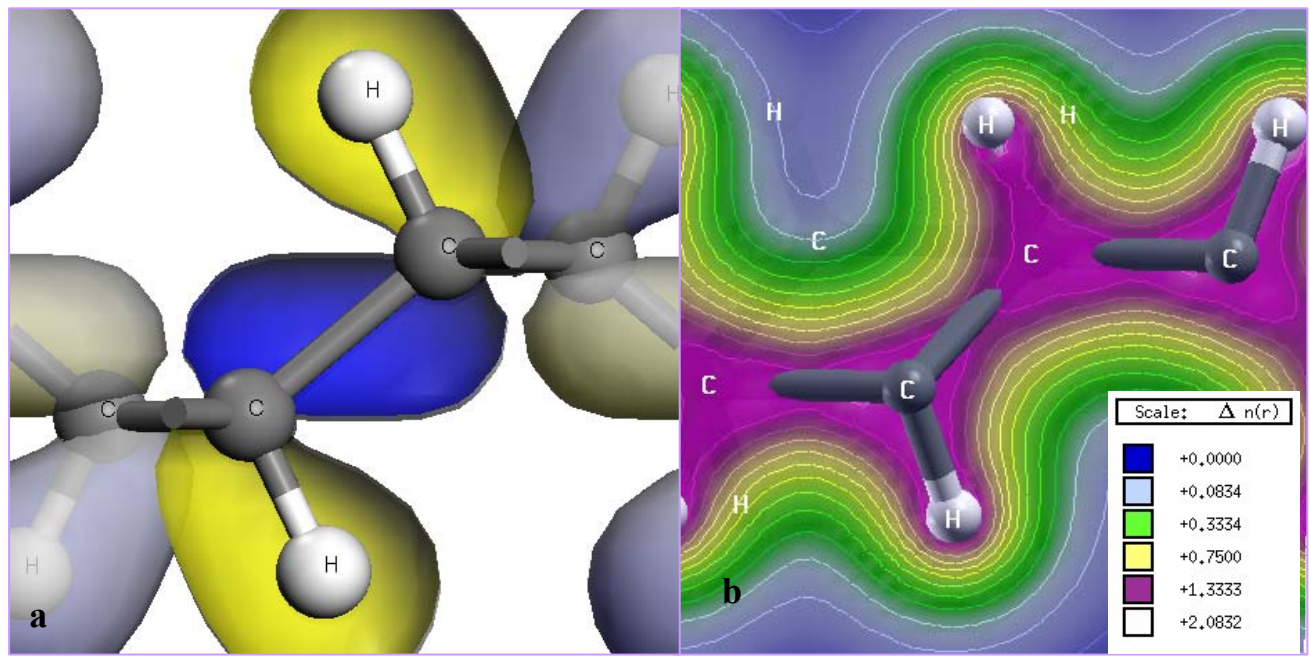

Fig 3.

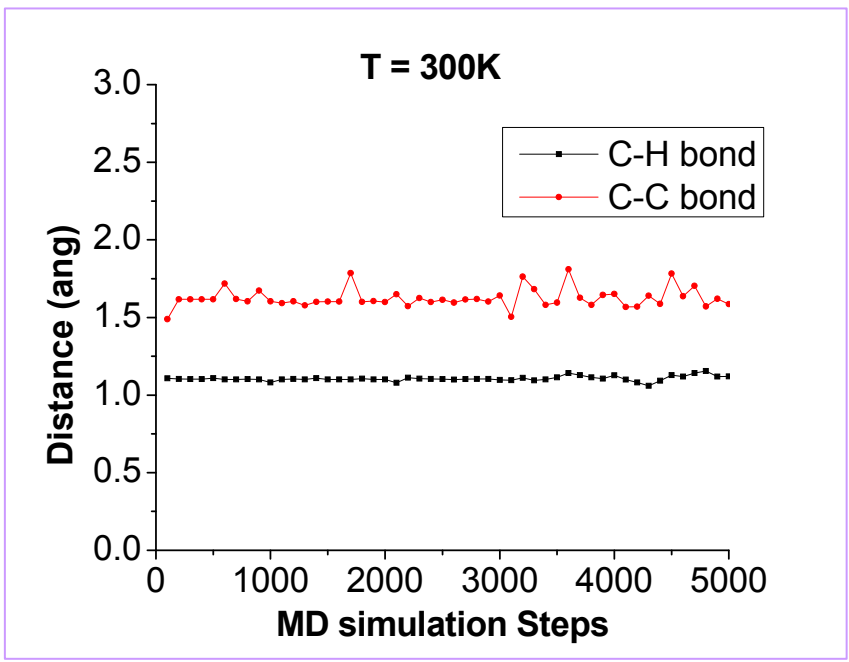

Fig 4.

Table1. Comparison of structure and ground state properties of the three conformers of graphane.

| Properties | | | Chair | Boat | Stirrup |
|---|---|---|---|---|---|
| Space group | | | P-3M1 (164) | PMMN (59) | PMNA (53) |
| Lattice Parameter (Å) | a | | 2.46 | 2.46 | 2.46 |
| | b | | 2.46 | 4.2608 | 10 |
| | c | | 10 | 10 | 4.2608 |
| Atomic positions | H | X | 0.666667 | 1.000000 | 1.000000 |
| | | Y | 0.333333 | 1.259391 | 0.651931 |
| | | Z | 0.352310 | 0.638576 | 0.470709 |
| | C | X | 0.666667 | 1.000000 | 1.000000 |
| | | Y | 0.333333 | 1.182504 | 0.551700 |
| | | Z | 0.456925 | 0.533077 | 0.358844 |
| Bond lengths (Å) | C-C | | 1.50 | 1.51, 1.56 | 1.53, 1.58 |
| | C-H | | 1.11 | 1.10 | 1.10 |
| Bond angle ∟HCC | | | 108º | 107º | 105º |
| BE (eV/ atom) | | | 5.21 | 5.18 | 5.16 |
| Band gap (eV) | | | 3.5 | 3.3 | 3.1 |
| Vibrational freq. of highest freq. mode (cm$^{-1}$) | | | 2969 | 3049 | 3122 |
| Mulliken charge transfer (electrons) | C | | -0.13 | -0.15 | -0.10 |
| | H | | +0.13 | +0.15 | +0.10 |